\documentclass[12pt,preprint]{aastex}

\usepackage{epsfig}
\usepackage{lscape}
\newcommand{\chisq}{$\chi^{2}$}
\newcommand{\G}{XGPS-I J183251-100106}
\newcommand{\XMM}{2XMM~J183251.4-100106}

\slugcomment{}

\shorttitle{XMM-Newton observation of \G}
\shortauthors{Hui et al.}

\begin{document}

\title{Detection of X-ray periodicity from a new eclipsing polar candidate \G} 

\author{C. Y. Hui\altaffilmark{1}, K. A. Seo\altaffilmark{1}, 
C. P. Hu\altaffilmark{2}, L. C. C. Lin\altaffilmark{3}, 
and Y. Chou\altaffilmark{2}}

\altaffiltext{1}{Department of Astronomy and Space Science, 
Chungnam National University, Daejeon 305-764, Korea}
\altaffiltext{2}{Graduate Institute of Astronomy, National Central University,
 Jhongli 32001, Taiwan}
\altaffiltext{3}{General Education Center, China Medical University, Taichung 40402, Taiwan}


\begin{abstract}
We report the results from a detailed analysis of
an archival XMM-Newton observation of the X-ray source \G, which has been suggested as 
a promising magnetic cataclysmic variable candidate based on its optical properties. 
A single periodic signal of $\sim1.5$~hrs is detected from all EPIC cameras on board XMM-Newton. 
The phase-averaged X-ray spectrum can be well-modeled with a thermal bremsstrahlung of a temperature 
$kT\sim50$~keV. Both X-ray spectral and temporal behavior of this system suggest it as a 
eclipsing cataclysmic variable of AM Herculis (or polar) type. 
\end{abstract}


\section{Introduction}
XMM-Newton Galactic plane survey (XGPS) has revealed a large population of X-ray sources (Hands et al. 2004).
In order to identify the nature of the brightest sources detected in XGPS, an optical campaign has recently
been carried out (Motch et al. 2010). In this process, three sources
were identified as promising cataclysmic variables (CVs).

The X-ray source designated as XGPS-9 (=\G) in Motch et al. (2010)
is one of the newly found CV candidates in XGPS.
Two optical objects are found within its X-ray error circle, while the fainter one ($V\sim23.3$)
is suggested to be the promising counterpart.
Its CV nature was identified through optical spectroscopy 
which unambiguously shows the strong H, He~I and He~II emission
along with a blue continuum (Motch et al. 2010). These features are known to be typical for a magnetic CV.

Apart from the optical identification, Motch et al. (2010) have also reported a brief X-ray analysis of
archival XMM-Newton data obtained by the observations on 15 March 2002 (Obs. IDs 0135741601, 0135744401).
The authors found that an absorbed
single component thermal plasma model is able to describe the data. However, as XGPS-9 located at an off-axis
angle $>9'$ in these short exposures ($\sim8$~ks and $\sim5$~ks respectively), the poor statistic 
and the degraded angular resolution leave
the spectral parameters unconstrained. While the X-ray light curve of
XGPS-9 (Fig.~8 in Motch et al. 2010) clearly shows the variability of this source, these
short observations preclude the search for any periodic behavior of this system.

In order to tightly constrain its X-ray properties, we have performed a follow-up investigation
of XGPS-9 with a deep XMM-Newton observation. The results of this analysis are presented in this paper.

\begin{figure}[t]
\centerline{\psfig{figure=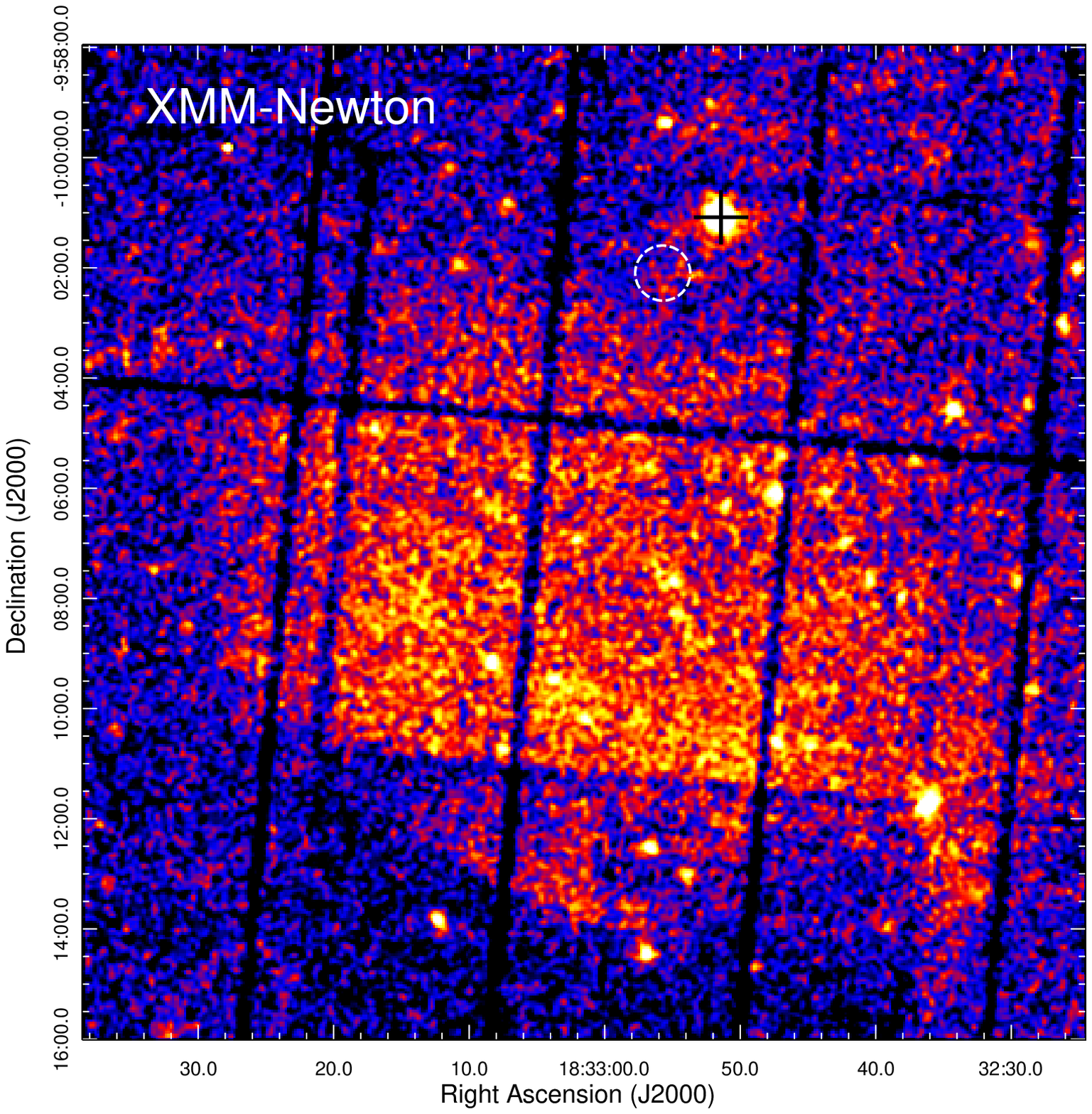,width=16cm,clip=}}
\caption[]{X-ray image of a $18'\times18$ FOV toward the supernova remnant Kes~69 with the MOS1/2 and PN data
merged. \G\ is the bright serendipitous source marked by a black cross. The dashed circle represents the 
background region adopted for the spectral analysis (see text).}
\label{xmm_img}
\end{figure}

\section{Observation \& Data Analysis} 
XGPS-9 was observed by the {\bf E}uropean {\bf P}hoton
{\bf I}maging {\bf C}amera (EPIC) on board XMM-Newton, which consists of
two Metal Oxide Semiconductor CCD detectors (MOS1 and MOS2) and a PN CCD detector, on 8-9 October 2009
(Obs. ID: 0605480101). While the MOS1/2 cameras were operated in Full Frame mode with a a temporal resolution of 2.6~s, 
the PN camera was operated in Extended Full Frame mode with
a temporal resolution of 199.1~ms. Medium filters were used to block optical stray light in all 
EPIC cameras. 

The aimpoint of this observation is RA=$18^{\rm h}32^{\rm m}57.01^{\rm s}$
Dec=$-10^\circ 05' 41.0''$ (J2000). 
Using tasks {\it emproc} and {\it epproc} of the XMM Science Analysis Software
(XMMSAS version 11.0.0), we have reprocessed all the EPIC data with the updated instrumental calibration. 
We subsequently selected 
only those events for which the pattern was between $0-12$ for both MOS cameras and $0-4$ for the PN
camera in $0.3-12$ keV. We further cleaned the data by accepting only the good times when sky background was low 
for the whole camera ($<2.4$~counts/s, $<2.8$~counts/s and $<5.7$~counts/s for MOS1, MOS2 and PN respectively.). 
After removing all events potentially contaminated by bad pixels, the effective
exposures are found to be 55.9~ks, 56.6~ks and 41.3~ks for MOS1, MOS2 and PN respectively.

We have also utilized the Optical Monitor (OM) data for our investigation. XGPS-9 has been observed by OM  
in standard imaging mode with three different filters: U (effective wavelength 3440\AA), UVW1 (2910\AA) and UVM2 (2310\AA). 
The exposures of OM with U, UVW1 and UVM2 are 4.4~ks, 4.4~ks and 2.5~ks respectively. 
For data reprocessing, flat-fielding and source detection, we have utilized the metatask {\it omichain}. 

This observation was originally intended for spectro-imaging investigation of the supernova remnant Kes~69
(see Figure~\ref{xmm_img}), which has its X-ray properties poorly constrained (cf. Yusef-Zadeh et al. 2003,
Bocchino et al. 2012).
A detailed analysis of the remnant emission from Kes~69 will be published elsewhere 
(Seo et al. in preparation) as this is out of scope in
this paper. In this observation, XGPS-9 is an serendipitous source located at an off-axis angle of $\sim5'$
(Figure~\ref{xmm_img}).
With the aid of the XMMSAS task {\it edetect\_chain}, we have run source detection individually on each 
EPIC data set. The mean X-ray position of XGPS-9 determined from this observation is 
RA=$18^{\rm h}32^{\rm m}51.51^{\rm s}$ Dec=$-10^\circ 01' 05.0''$ (J2000) with a resultant uncertainty of 
$0.44''$ by combining the statistical errors inferred from each camera in quadrature. 
This X-ray position is marked as a black cross in Figure~\ref{xmm_img}.

For timing analysis, the point source was extracted within a circular region of a 
$15''$ radius centered at the X-ray position from each camera, which corresponds to an encircled energy fraction
of $\sim70\%$. 758 counts, 682 counts and 1277 counts were obtained from MOS1, MOS2 and PN respectively.
All the photon arrival times are subsequently corrected to the solar system barycenter with the
XMMSAS tool {\it barycen} by adopting the updated ephemeris JPL DE405\footnote{We note that in comparing with 
DE405, the previous generation of planetary ephemeris, DE200, has a $\sim300$~m error in Earth's position. This 
corresponds to a timing error of $\sim1$~$\mu$s which is negligible in our investigation of orbital period.} 
and the aforementioned mean X-ray position.

\begin{figure}[t]
\centerline{\epsfig{figure=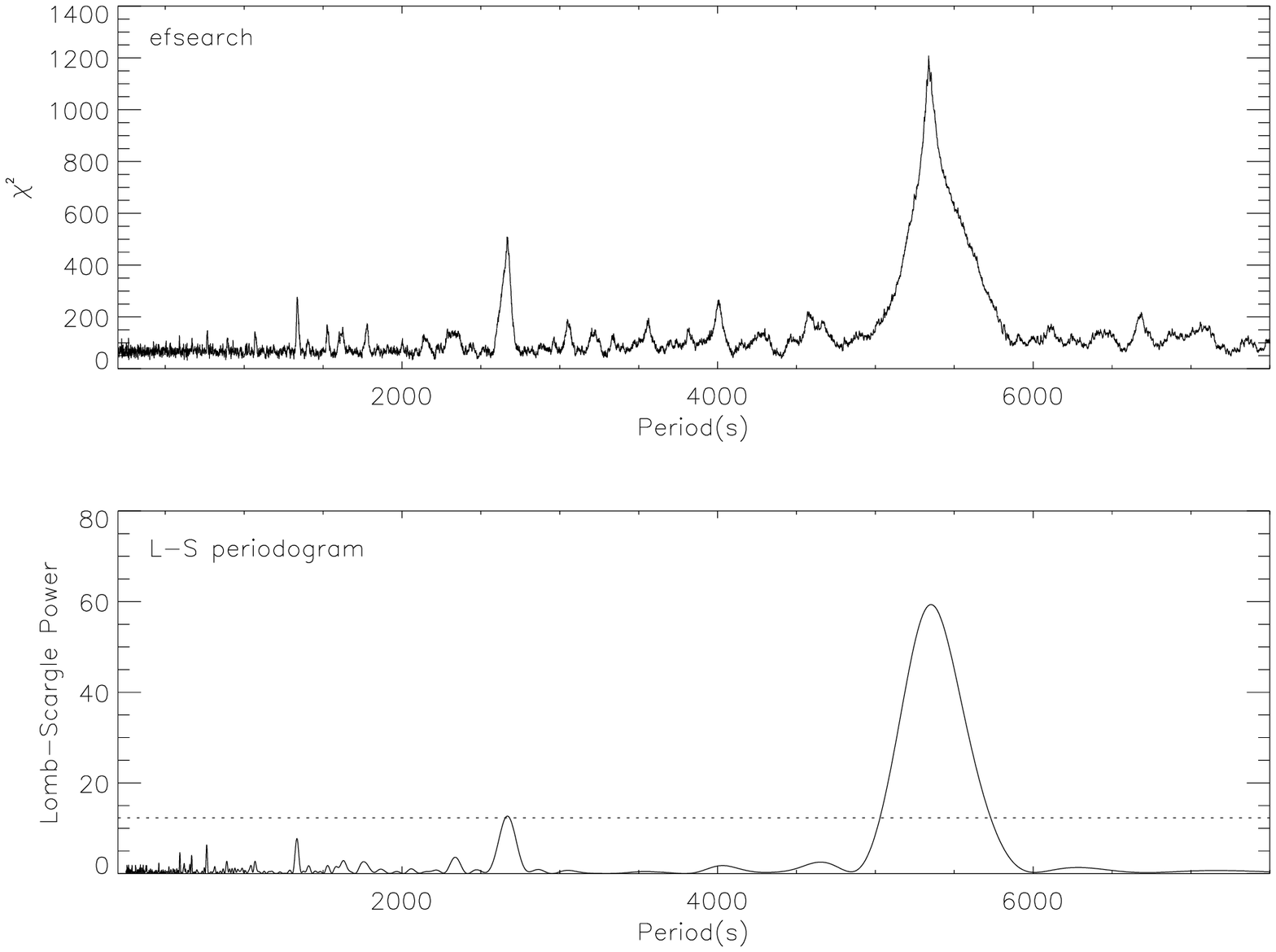,width=16cm,clip=}}
\caption[]{The result of epoch-folding period search (upper) and the Lomb-Scargle 
periodogram (lower). The dashed line in the Lomb-Scargle periodogram is the 3$\sigma$ significance level.}
\label{efsearch_ls}
\end{figure}

In order to search for any periodicity, we have applied the techniques of both epoch-folding 
and the Lomb-Scargle periodogram (Scargle 1982) on the event list and the binned light curve respectively.
The epoch-folding method was performed with the \emph{efsearch} subpackage of HEAsoft.
The peak value of the $\chi^2$ spectrum with 50 bins for the arrival time
of merged MOS and PN events is at $P=5337.7\pm21.7$s.
The uncertainty of epoch-folding is estimated according to the empirical formula of Leahy (1987).
On the other hand, we have binned all of the events into a light curve with 150~s resolution to 
calculate the Lomb-Scargle periodogram.
The periodogram demonstrates a periodicity at $P=5353.0\pm13.5$s which is consistent 
with the result of epoch-folding. 
The uncertainty was estimated by $10^4$ times of Monte Carlo simulation.
The results of the periodicity search with epoch-folding and
Lomb-Scargle method are shown in Fig.~\ref{efsearch_ls}.

\begin{figure}[t]
\centerline{\epsfig{figure=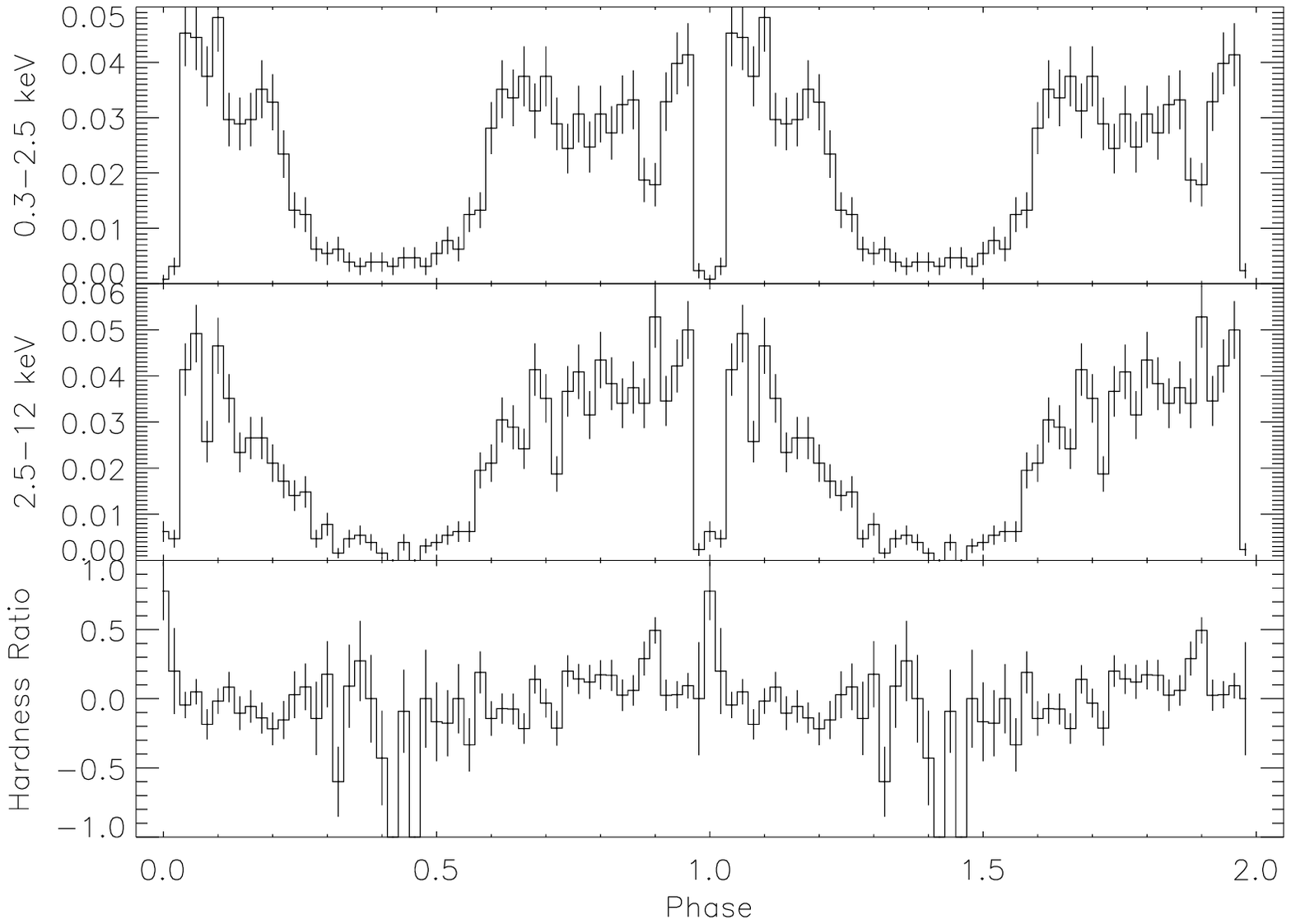,width=16cm,clip=}}
\caption[]{The folded light curve of soft band (upper), hard band (middle), and the corresponding hardness ratio (bottom). 
The unit of $y-$axis for the first two panels is count~s$^{-1}$. We have adopted a resolution of 50 bins per cycle.} 
\label{efold_hardness}
\end{figure}

In order to identify the nature of this periodicity, we have also folded the photon events
based on the period we detected. The 5337.7s period was adopted to fold the light curve as it is directly 
resulted from epoch-folding. In addition, we have divided the photons into soft band (0.3-2.5~keV) 
and hard band (2.5-12~keV) for computing the hardness ratio: (hard-soft)/(hard+soft).
The folded light curves of both bands and the hardness ratio are shown in
Fig.~\ref{efold_hardness}, where the epoch zero is set at MJD~55113.0383.

The minimum at phase zero is particularly interesting and leads us to further investigate it. 
This feauture resembles those resulted from the eclipses of X-ray emitting regions in polars 
(e.g. Mukai et al. 2003; Vogel et al. 2008). We notice there is residual X-ray emission at the minimum.  
To investigate if the residual is caused by the background, we estimated the count rates within a number of 
source-free circular regions of 15" radius around XGPS-9. The average background count rate 
estimated from the merged dataset is 
$(3\pm2)\times10^{-3}$~counts/s which is consistent with the minimum of the folded light curve in 0.3-12~keV. 
This suggests the residual X-rays are likely the background contamination. Therefore, 
the feature can possibly be a total eclipse of the X-ray emitting region by the companion star.

To further characterize this feature, we attempted to estimate the eclipse width and to obtain an ephemeris. 
In view of the relatively low statistics, these properties cannot be properly constrained by fitting the 
eclipse profile. Instead, visual inspection was adopted for the measurements reported in this paper. 
The folded light curves shown in Figure~\ref{efold_hardness} suggests the ingress and egress 
are rather sharp which allow us to estimate the eclipse duration with the flat bottom minimum. This leads to 
an approximate width of $320\pm53$~s, where the uncertainty is estimated from half of the bin size in 
Figure~\ref{efold_hardness}. 

For defining the mid-eclipse phase, we adopted the mid-point of this feature in Figure~\ref{efold_hardness}. 
The timing of the X-ray eclipse can be described by the following linear ephemeris:

\begin{equation}
T_{N(\rm mid-eclipse)}={\rm MJD(TDB)}~\left(55113.0383\pm6.2\times10^{-4}\right)+\frac{5337.7\pm21.7}{86400}\times N.
\end{equation}

We have also checked these results by adopting different bin size (32, 64, 128 bin/period), 
we found that both the eclipse width and the 
ephemeris are consistent with the aforementioned results within the tolerence of uncertainties. 

To investigate if the periodic signal comes from the synchronous rotation, we further
examined the existence of any marginal periodic signal in a larger frequency range using the
$\chi^2$-test with 32 bins and Fourier resolution.
Because the spin period of a typical white dwarf is about hundreds of seconds (cf. Ritter \& Kolb 2003) 
and the signal
of orbital modulation seriously contaminates the periodogram when the searching range is larger
than 2.5$\times 10^{-3}$~s$^{-1}$, we only considered a periodic signal of spin within 100--400~s.
Within this selected range, the largest peak value can be found at 164.7(1)~s with the chance probability
of $9.6 \times 10^{-5}$. However, this result is not significant if we took into account the number
of trials in the searching range.
We also tried the decomposition methods (e.g. the empirical mode decomposition proposed
by Huang et al. 1998) to resolve another strong signal.
However, we did not yield any other possible periodicity except for the binary modulation.
Therefore, we conclude that no spin period can be detected from the current
{\it XMM-Newton} observation and it may give a hint of the synchronized spin and orbital periods.

Motch et al. (2010) have briefly reported the X-ray spectral properties of XGPS-9. They found that a
one-component thermal plasma spectrum with a column absorption of $N_{H}\sim8\times10^{21}$~cm$^{-2}$
can provide a reasonable fit to the data. However, due to the limited photon statistic, the plasma temperature
and the X-ray flux in their study remains to be unconstrained. With a much longer integration time of the data 
adopted in our investigation, we are able to provide a tighter constraint on its spectral 
properties. For the spectral analysis, we have extracted the spectra within a circular 
region with a $30''$ radius in each MOS camera and a $25''$ radius in PN camera around the X-ray position 
respectively. This corresponds to an encircled energy fraction of $\sim85\%$ in all EPIC cameras. 
Spectra extracted 
from each camera were grouped so as to have at least 50 counts per spectral bin.
The background spectra were 
sampled from the low count circular region of $30''$ radius centered at
RA=$18^{\rm h}32^{\rm m}55.72^{\rm s}$, Dec=$-10^{\circ}02^{'}05.87^{''}$ (J2000) in each camera. 
Response files were computed by using the 
XMMSAS tasks \emph{rmfgen} and \emph{arfgen}. For the spectral analysis, we used XSPEC (version 12.6.0) 
with $\chi^{2}$ statistics adopted for all the fittings. All the quoted uncertainties of the spectral parameters 
are $1\sigma$ for 1 parameter of interest.

\begin{figure*}[t]
\centerline{\psfig{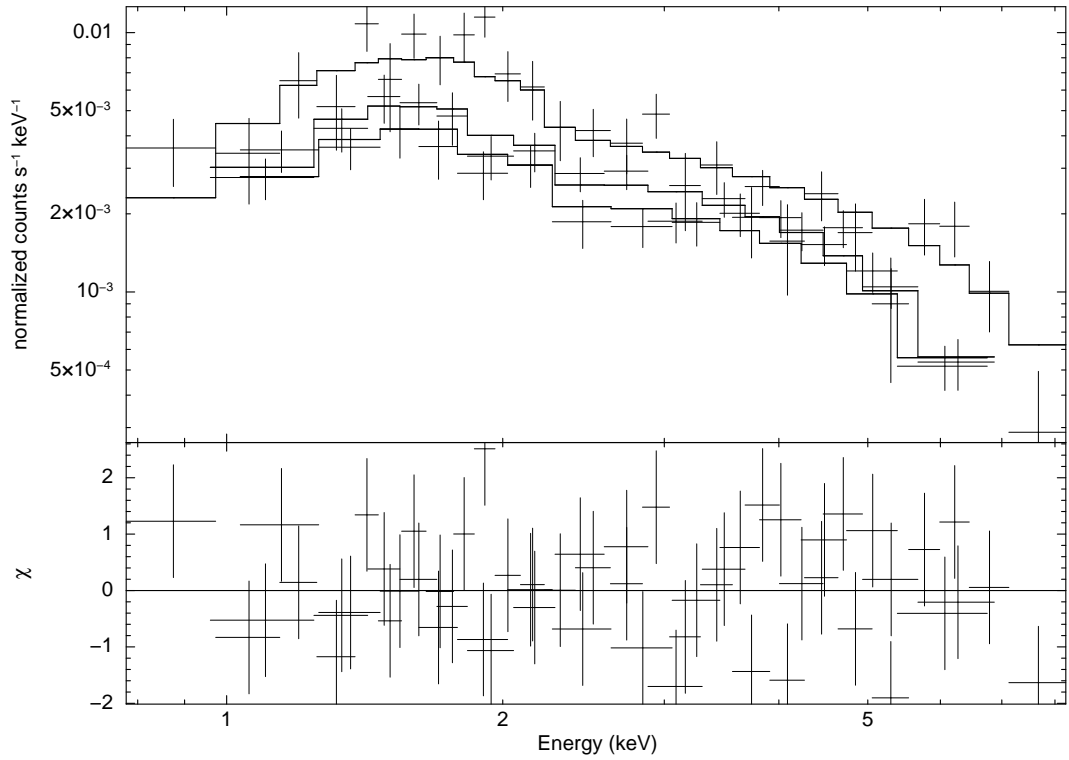}}
\caption[]{Energy spectrum of \G\ as observed with the PN (upper spectrum)
and MOS1/2 detectors (lower spectra) and simultaneously fitted to an absorbed
thermal bremsstrahlung model
({\it upper panel}) and contribution to the \chisq\, fit statistic
({\it lower panel}).}
\label{xmm_spec}
\end{figure*}

\begin{figure}[t]
\centerline{\psfig{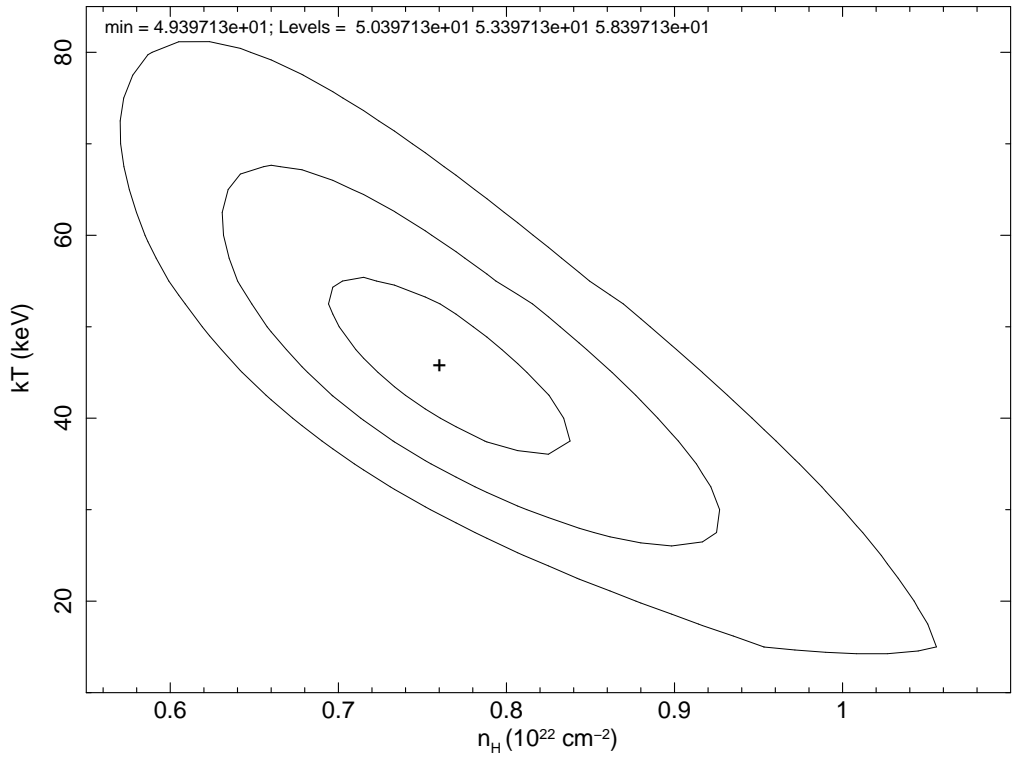}}
\caption[]{$1\sigma$, $2\sigma$ and $3\sigma$ confidence contours for the thermal bremsstrahlung 
model fitted to the X-ray spectrum of \G. The $\chi^{2}$ values for the minimum and each contour are 
given at the top of the figure.}
\label{contour}
\end{figure}

Based on the optical properties, Motch et al. (2010) suggested XGPS-9 a magnetic CV. To explore this scenario,
we examine its X-ray spectrum by fitting an absorbed thermal bremsstrahlung model, which accounts for 
the X-ray emission from the shock-heated gas in the accretion column. We found that this single component 
model can describe the data reasonably well with a goodness-of-fit of \chisq=49.40 for 54 d.o.f.. The spectral fit 
yields a column density of $N_{\rm H}=7.6^{+0.8}_{-0.7}\times10^{21}$~cm$^{-2}$, a plasma temperature of 
$kT=46\pm10$~keV and 
an unabsorbed flux of $f_{\rm x}=3.3^{+0.9}_{-0.2}\times10^{-13}$~erg~cm$^{-2}$~s$^{-1}$ 
in the energy range of $0.3-10$~keV. The 
comparison between the best-fit model and the data is shown in Figure~\ref{xmm_spec}. 
We have also computed the confidence contours to investigate the relative parameter 
dependence between the plasma temperature and the column absorption, which are plotted in 
Figure~\ref{contour}.

As X-ray spectrum of CVs can possibly show the presence of line emission, we have also examined the 
observed spectrum of XGPS-9 with XSPEC model MEKAL which is a code that models the plasma in 
collisional ionization equilibrium with the line emission built in. We found that the fit yields 
a comparable goodness-of-fit with the thermal bremsstrahlung model (\chisq=50.11 for 54 d.o.f.). 
The best-fit spectral parameters are also similar ($N_{\rm H}=7.5\pm0.7\times10^{21}$~cm$^{-2}$, 
$kT=50^{+13}_{-11}$~keV). 


For magnetic CVs, part of the thermal bremsstrahlung emission can possibly be Compton reflected by the 
white dwarf surface (see Matt et al. 1998). This so-called Compton reflection 
component is expected to be accompanied with a fluorescent K$\alpha$ iron line (de Martino et al.2008). 
For searching evidence for the line feature, we have added a Gaussian component to the thermal 
bremsstrahlung model and fixed the line energy at 6.4~keV. To begin with, we have allowed both normalization 
and the line width as free parameters. However, we found that the line width cannot be constrained (i.e. 
the fitted width is essentially zero). 
This led us to fix it at a reasonable value of $\sigma=100$~eV (see Balman 2011; de Martino et al. 2008). 
This results in a best-fit parameter set of $N_{\rm H}=7.8^{+0.9}_{-0.8}\times10^{21}$~cm$^{-2}$, 
$kT=36\pm10$~keV, and a line flux of $f_{\rm 6.4~keV}=3.0^{+4.5}_{-3.0}\times10^{-7}$~photons~cm$^{-2}$~s$^{-1}$
with \chisq=49.11 for 53 d.o.f.. In view of the large uncertainty of fitted line flux and 
the insignificant improvement of the goodness-of-fit, we conclude that there is no compelling evidence of 
Fe~K$\alpha$ line can be found in this observation. We report a $1\sigma$ upper limit on the line flux of 
$f_{\rm 6.4~keV}<7.5\times10^{-7}$~photons~cm$^{-2}$~s$^{-1}$. 

While a thermal bremsstrahlung/MEKAL can model the spectrum well, as mentioned by Motch et al. (2010), the spectral 
hardening can possibly be resulted from the reflection component and/or warm absorption 
(see Schwarz et al. 2009; Staude et al. 2008). Therefore, the 
plasma temperature inferred from this simple model can be unphysically high.
We have also investigated if any other single 
component model can provide any acceptable phenomenological description. We notice that a simple 
absorbed power-law can also result 
in a comparable goodness-of-fit (\chisq=49.83 for 54 d.o.f.), and hence we cannot distinguish it from 
the thermal bremsstrahlung/MEKAL model statistically. The power-law spectral fit yields 
$N_{\rm H}=7.9^{+1.3}_{-0.9}\times10^{21}$~cm$^{-2}$, a photon index of $\Gamma=1.3\pm0.1$ and an 
unabsorbed flux of $f_{\rm x}=3.4^{+0.9}_{-0.7}\times10^{-13}$~erg~cm$^{-2}$~s$^{-1}$ ($0.3-10$~keV). 
The spectral steepness and the flux are fully consistent with those inferred from a brief analysis 
limited to the hard band (i.e. $3-10$~keV) as reported recently (cf. Tab.~1 in Bocchino et al. 2012). 

We have checked the robustness of all the spectral results quoted in this paper by incorporating background spectra 
sampled from different source-free regions around XGPS-9. We found that the spectral parameters inferred from the fittings 
with different adopted background are consistent within $1\sigma$ uncertainties. 

\begin{figure}[t]
\centerline{\epsfig{figure=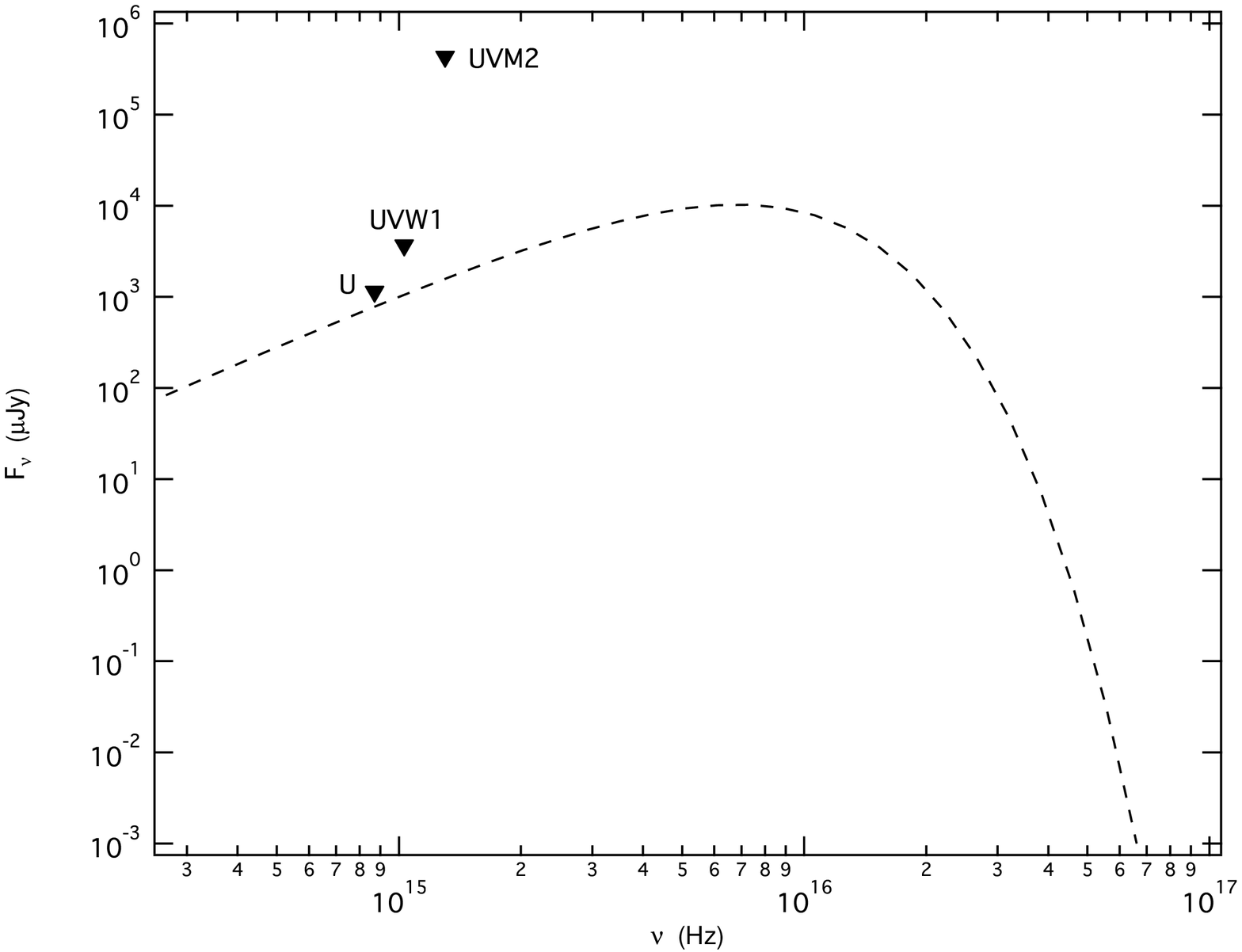,width=16cm,clip=}}
\caption[]{The dash line represents the upper-bound of 
possible blackbody contribution at $kT=10$~eV as constrained by the extinction-corrected limiting flux 
densities deduced 
from the OM data (triangular symbols).}
\label{om_sed}
\end{figure}

We have also attempted to constrain the possible contribution from the soft component by using OM data.
Soft component from a CV can be originated from the optically-thick blackbody like emission of the white dwarf 
surface. The temperature can range from few tens to hundred of eV (Evans \& Hellier 2007). Searching for 
the optical/UV counterpart in OM data results in non-detection in all three bands, which place a limiting 
flux density of $<2.3~\mu$Jy, $<2.7~\mu$Jy and $<7.8~\mu$Jy in U, UVW1 and UVM2 respectively. 
We adopted the column density inferred from the X-ray spectral fitting (i.e. $7.5\times10^{21}$~cm$^{-2}$) 
to perform the extinction-correction (cf. Predehl \& Schmitt 1995; Cardelli et al. 1989). With the de-reddened 
limiting flux densities, we place an upper bound for any soft component contribution by assuming a low temperature 
white dwarf with $kT=10$~eV (see Figure~\ref{om_sed}). The blackbody gives a limiting flux of 
$<5.4\times10^{-18}$~erg~cm$^{-2}$~s$^{-1}$ in $0.3-10$~keV, which is $<0.002\%$ of the total X-ray flux 
in this energy range.

\section{Summary \& Discussion}
Utilizing the archival {\it XMM-Newton} data, we have detected the periodicity from a new CV candidate XGPS-9. 
Through a detailed timing analysis, only a single periodic signal of $\sim1.5$~hrs can be found. 
This suggests a possible scenario of synchronous rotation, which favors the interpretation 
that the system is a magnetic CV of polar type. 

For a polar, the white dwarf is accreting matter from its late-type main sequence companion via 
Roche-lobe overflow which will be driven by the strong magnetic field of the white dwarf (with a 
surface magnetic field strength of few tens of million Gauss) to form a quasi-radial flow 
towards the magnetic pole. The accretion flow will eventually form a strong shock above the 
stellar surface. The shock-heated plasma will be cooled via bremsstrahlung radiation in X-ray regime with 
typical temperature of few tens of keV, which is consistent with the best-fit plasma temperature 
(i.e. $kT=46\pm10$~keV) inferred in our spectral analysis. 

It is interesting to compare the temporal behavior of XGPS-9 and 2XMMp~J131223.4+173659, 
which is a eclipsing polar serendipitously identified in 2XMM catalog (Vogel et al. 2008). 
The properties of these two systems are remarkably similar. First, both objects show a 
narrow and deep minimum in their phase-folded light curves (i.e. depression in phase $0.97-1.03$ 
for the case of XGPS-9). This feature is interpreted as the eclipse of X-ray emission 
in the polar region of the white dwarf when the companion star passes through our line-of-sight towards 
this region. Also, the broad depression (i.e. phase $\sim0.2-0.6$ in 
Fig.~\ref{efold_hardness}), which is likely due to the self-occultation by the white dwarf itself, is 
observed in both system. 
The coexistence of these two features indicates that the spin of the white dwarf
and the binary orbital periods are synchronized.
Furthermore, a shallow minimum in the soft band light curve (in phase $\sim0.7-0.9$ in Fig.~\ref{efold_hardness}) 
is found before the eclipse. 
When the X-ray emitting region is obscured by the accretion stream, a pre-eclipse dip can be resulted from 
the increased photoelectric absorption which makes the dip more prominent in the soft band 
(see Fig.~\ref{efold_hardness}). 

We further attempted to place a constraint on the orbital inclination with our estimated eclipse width 
(i.e. $\sim320$~s). Assuming the companion star fills its Roche lobe and X-ray emitting region is close to 
the surface of the white dwarf, the duration of the total eclipse of the white dwarf can be expressed as a function 
of mass ratio and the inclination (cf. Fig.~2 in Horne 1985). The eclipse width corresponds to a phase interval 
of $\Delta\phi\sim0.06$ which we considered as an upper limit of the eclipse width as a longer observation 
might reveal the falling/rising edges in the eclipse profile. We further assume that the companion star is 
a M-dwarf with mass 
$M_{*}\gtrsim0.1~M_{\odot}$. For a typical white dwarf (i.e. $M_{\rm wd}\sim0.6~M_{\odot}$), this implies a 
mass ratio of $q=\frac{M_{*}}{M_{\rm wd}}\gtrsim0.17$. With these assumptions, the upper bound of the orbital 
inclination can be placed at $\lesssim80^{\circ}$. 

For further investigation of XGPS-9, dedicated optical/infrared observations will be useful for confirming 
its source nature. A differential photometric study can enable a comparison with the X-ray light curve reported 
in this paper. This can provide information for further constraining the accretion geometry. Also, the free electrons 
in the shock region can spiral in the magnetized ionized gas. This can lead to the emission of cyclotron radiation 
in infrared regime. Therefore, infrared spectroscopy and polarimetry are encouraged for determining the magnetic 
field of the white dwarf directly. We would also like to point out that the current data 
cannot provide a conclusive constraint on the plasma temperature of a thermal bremsstrahlung spectrum at a few tens of keV, 
which is far above the energy coverage of XMM-Newton.  
This is reflected by its large relative uncertainty reported in this work (i.e. $\gtrsim20\%$). 
For a more constraining spectral results, hard X-ray 
observation is certainly needed. As XGPS-9 resides in rather complex region, the recently commenced mission 
NuSTAR and the upcoming one (e.g. Astro-H), which equipped with hard X-ray focusing optics, will provide the 
desirable instruments for the further investigations of this interesting eclipsing polar.

\acknowledgments{
CYH and KAS are supported by the National Research Foundation of Korea through grant 2011-0023383. LCCL 
is supported by the National Science Council of Taiwan through the grant NSC 101-2112-M-039-001-MY3. 
CPH and YC are supported by the National Science Council of Taiwan through the grant 
NSC 101-2112-M-008-010. 
}


\end{document}